\documentclass[12pt]{iopart}

\usepackage{subfigure}
\usepackage{graphicx}
\begin{document}

\title[]{Exact $P_s T_d$ Invariant and $P_s T_d$ Symmetric Breaking solutions, Symmetry Reductions and B\"{a}cklund Transformations For An AB-KdV System}

\author{Man JIA$^{1,2}$, Sen Yue LOU$^{1,2,3}$}
\address{$^1$ Physics Department, Ningbo University, Ningbo 315211, China}
\address{$^2$ Ningbo Collaborative Innovation Center of Nonlinear Hazard System of Ocean and Atmosphere, Ningbo University, Ningbo 315211, China}
\address{$^3$ Shanghai Key Laboratory of Trustworthy Computing, East China Normal University, Shanghai 200062, China}

\ead{jiaman@nbu.edu.cn}
\vspace{10pt}

\begin{abstract}
In natural and social science, many events happened at different space-times may be closely correlated. Two events, A (Alice) and B (Bob) are defined as correlated if one event is determined by another, say, $B=\hat{f} A$ for suitable $\hat{f}$ operators. A nonlocal AB-KdV system with shifted-parity ($P_s$, parity with a shift), delayed time reversal ($T_d$, time reversal with a delay) symmetry where $B=\hat{P_s}\hat{T_d} A$ is constructed directly from the normal KdV equation to describe two-area physical event. The exact solutions of the AB-KdV system, including $P_s T_d$ invariant and $P_s T_d$ symmetric breaking solutions are shown by different methods. The $P_s T_d$ invariant solution show that the event happened at $A$ will happen also at $B$. These solutions, such as single soliton solutions, infinitely many singular soliton solutions, soliton-cnoidal wave interaction solutions, and symmetry reduction solutions etc., show the AB-KdV system possesses rich structures. Also, a special B\"{a}cklund transformation related to residual symmetry is presented via the localization of the residual symmetry to find interaction solutions between the solitons and other types of the AB-KdV system.
\end{abstract}

\pacs{05.45.Yv, 02.30.Ik, 02.30.Jr}
%
\vspace{2pc}
\noindent{\it Keywords}: AB-KdV system, $P_s T_d$ symmetry, $P_s T_d$ invariant solutions, $P_s T_d$ symmetric breaking solutions, B\"{a}cklund transformation
%
%
%
%

\section{Introduction}
2013, Mark J. Ablowitz \cite{ablowitzprl} introduced a nonlinear Schr\"{o}dinger equation
\begin{eqnarray}
i q_t(x, t)=q_{x x}(x, t) \pm 2 q(x, t)q^*(-x, t) q(x, t),\label{nls}
\end{eqnarray}
where $*$ denotes complex conjugation and $q(x, t)$ is a complex valued function of the real variables $x$ and $t$. Some of the important properties of the nonlocal NLS equation (\ref{nls}) were contrasted with the classical NLS equation where $q^*(-x, t)$ is replaced by $q^*(x, t)$. The author believed both eq. (\ref{nls}) and the classical NLS share the symmetry that when $x \rightarrow -x$, $t\rightarrow -t$ and a complex conjugate is taken, the equation remains invariant.

Note that eq. (\ref{nls}) is invariant neither under parity $P$, whose effect is to make spatial reflections, $p \rightarrow -p$ and $x \rightarrow -x$, nor under time reversal $T$, which replaces $p \rightarrow -p$, $x \rightarrow -x$, and $i \rightarrow -i$. Similarly, other types of nonlocal nonlinear systems possessing $PT$ symmetry are constructed, such as the coupled nonlocal NLS systems \cite{songxiaozhu}, the nonlocal KdV and modified KdV systems \cite{new} - \cite{jizhu}, the discrete nonlocal NLS system \cite{ablowitzpre}, and the nonlocal Davey-Stewartson system \cite{dimakosjmp} - \cite{fokasnonli}, etc. Studies show that $PT$ symmetry is important in not only particle physics and quantum physics \cite{bender}, but also many other research fields in physics, such as optics \cite{makris}, quantum field theory \cite{bender2}, electric circuits \cite{lin}. Recently, $PT$-symmetric nonlocal NLS equation (\ref{nls}) is used to describe the extension of properties of traditional macroscopic magnetic systems \cite{gadzphysreva}.

In order to explore and solve the nonlocal nonlinear systems possessing $PT$ symmetry, many approaches are proposed, such as the inverse scatting method \cite{ablowitzprl}, Darboux transformations and Backlund transformations \cite{songxiaozhu}. Many interesting phenomenon have been reported about the systems that possesses $PT$ symmetry.

There are other possible crucial properties in physics, such as charge conjugation ($C$), shifted-parity ($P_s$, parity with a shift, or $x \rightarrow -x+x_0$, where $x_0$ is an arbitrary constant), delayed time reversal ($T_d$, time reversal with a delay, or $t \rightarrow -t+t_0$, where $t_0$ is an arbitrary constant) and their possible combinations such as $PT$, $PC$, $P_s C$, $P_s T_d$ which can be successively used to describe two-place physical problems. These systems are named Alice-Bob (AB) systems \cite{lou1} introduced by the AB-BA principle and $P_s-T_d-C$ principle. If one event ($A$, Alice event) is correlated/entangled to another ($B$, Bob event), we denote the correlated relation as $B=\hat{f} A$ for suitable $\hat{f}$ operators, such as $PT$, $PC$, $P_s C$, $P_s T_d$. Usually, event $A=A(x,\ t)$ happened at $\{x,\ t\}$ and event $B=B(x',\ t')$ happened at $\{x',\ t'\}=\hat{f}\{x,\ t\}$. In fact, $\{x',\ t'\}$ is usually far away from $\{x,\ t\}$. Hence, the intrinsic two-place models or Alice-Bob systems are nonlocal. The nonlocal NLS equation (\ref{nls}) can be considered as a special type of the AB systems.

As the widely used PT symmetry, considered as special type of AB systems, in particle physics and quantum physics, AB systems also play an important role in natural and social science for many events happened at different space-times may be closely correlated. For instance, in quantum physics, many faraway particles (atoms) may construct a completely entangled state, a measurement of one particle (event $A$) will affect the state of the other (event $B=\hat{f} A$). Recent studies show a special AB-KdV system is derived from the multiple vorticity interaction model which is related to a standard atmospheric and oceanic dynamic system, the nonlinear inviscid dissipative and barotropic vorticity equation in a $\beta$-plane channel in \cite{lou2}. It is found that the $P_sT_d$ symmetry breaking soliton solution of the derived AB-KdV system can be used to qualitatively describe the two real events, the atmospheric blocking happened in November $2007$ and January $2008$ respectively while the atmospheric blockings are responsible for the heavy snow disaster in Southern China in the winter $2007/2008$.

Many special types of AB systems such as the KdV-KP-Toda type, mKdV-sG type, NLS type and discrete $H_1$ type are established from the well-known integrable systems in \cite{lou1}. Some special $P_s-T_d-C$ group invariant multi-soliton solutions of the AB systems are obtained by using the $P_s-T_d-C$ principle. Furthermore, many types of $P_sT_d$ invariant solutions of a general AB-KdV system, such as the Painlev\'{e} II reduction and soliton-cnoidal periodic wave interaction solutions are constructed in \cite{lou2}. It is pointed out that the physical meaning of the $P_sT_d$ invariant solutions is that the event happened at $\{x,\ t\}$ will happen also at $\{x',\ t'\}$.

Except for the $P_sT_d$ invariant solutions, there are other types of multiple soliton solutions of $P_sT_d$ symmetry breaking for different AB models. The physical meaning of the group symmetry breaking solution is that
for Alice-Bob systems there are real physical phenomena where the event $B$ at $\{x',\ t'\}$ is different from the event $A$ at $\{x',\ t'\}$. A special $P_sT_d$ symmetry breaking solution for AB-KdV system is also listed in \cite{lou2}. However, to find $P_sT_d$ symmetry breaking solution is still more difficult.

Hence, more quite relevant and important questions arise: do there exist more possible AB systems that can be used to describe two-place event? Can we find out the exact solutions of the systems? How can we find the solutions? Can we find $PT$ symmetry solutions and $PT$ symmetry breaking solutions for the systems?

In this manuscript, we first construct a special AB-KdV system with the $P_sT_d$ principle, then try to find out the exact solutions, including $P_s T_d$ invariant and $P_s T_d$ symmetric breaking solutions of the AB-KdV system. The $P_s T_d$ invariant solutions are obtained by using the $P_s-T_d$ principle, while the $P_s T_d$ symmetric breaking solutions are solved by the help of a coupled KdV system. Moreover, in order to find interaction solutions between the solitons and other types of solitary waves, we also present the B\"{a}cklund transformation related to residual symmetry via the localization of the residual symmetry of the AB-KdV system.

\section{An AB-KdV System}
In this manuscript, we mainly discuss the AB-KdV system
\begin{eqnarray}
&& A_t+A_{xxx}+3(A+B)(B_x+3 A_x)=0, \nonumber \\ &&  B=\hat{P_s}\hat{T_d}A=A(-x+x_0,-t+t_0). \label{abkdv}
\end{eqnarray}
The AB-KdV system eq. (\ref{abkdv}) is established as follows according to \cite{lou2}. Substituting the ansatz $u= \frac{1}{2}(A+B)$ where $u=u(x,t)$ and $B=\hat{P_s}\hat{T_d}A=A(-x+x_0,-t+t_0)$ into the common KdV equation
\begin{eqnarray}
u_t+24 u u_x +u_{xxx}=0, \label{kdv}
\end{eqnarray}
we obtain an equation related to $A$ and $B$
\begin{eqnarray}
&& A_t+B_t+A_{xxx}+B_{xxx}+12(A+B)(A_ x +B_x)=0.\label{pstdsym}
\end{eqnarray}
It is easy to find out that the system (\ref{pstdsym}) can be split into two equations
\begin{eqnarray}
&&A_t+A_{xxx}+3(A+B)(B_x+3 A_x)+G(A, \ B)=0, \label{pi} \\ && B_t+B_{xxx}+3(A+B)(A_x+3 B_x)-G(A, \ B)=0, \label{pii}
\end{eqnarray}
where $G(A, \ B)$ may be an arbitrary function of $A$ and $B$. Applying the operator $P_sT_d$ on eq. (\ref{pii}), we  can see it will lead to a compatibility condition
\begin{eqnarray}
G(A, \ B)=\hat{P_s}\hat{T_d}G(A, \ B),\nonumber
\end{eqnarray}
which means the arbitrary function $G(A, \ B)$ should be $P_sT_d$ invariant. Generally there are numerous functions satisfying $P_s T_d$ invariant, but for simplicity, we only take $G(A, \ B)$ as zero. Then the equations (\ref{pi}) - (\ref{pii}) are reduced to the special AB-kdV system (\ref{abkdv}). As B is related to A by $B=A(-x+x_0,\ -t+t_0)$ which is far away from A, the AB-KdV system (\ref{abkdv}) is nonlocal.

Now the AB-KdV system (\ref{abkdv}) has been established, next we will search for the exact solutions of the AB-KdV system. As the system is nonlocal, it is still difficult to find out the exact solutions. Fortunately, the AB-KdV system (\ref{abkdv}) is directly constructed from the common KdV equation (\ref{kdv}) by using the $P_sT_d$ principle, so it is natural to construct the solutions of the AB-KdV system from the original KdV equation by selecting out the $P_s T_d$ solutions and rewriting them to satisfy the AB-KdV system. Thus the solutions of the AB-KdV system obtained by this method are invariant under the transformation of $x \rightarrow -x+x_0, t \rightarrow -t+t_0$, the solutions are $P_s T_d$ invariant. In the next section, we will present the $P_s T_d$ invariant solutions of the AB-KdV system (\ref{abkdv}) directly from the common KdV equation (\ref{kdv}).

\section{$P_s T_d$ invariant solutions to the AB-KdV system}
As the AB-KdV system (\ref{abkdv}) is directly derived from the common KdV equation (\ref{kdv}), all the $P_sT_d$ invariant solutions of the KdV equation (\ref{kdv}) are also solutions to the AB-KdV system (\ref{abkdv}). Various exact solutions of the KdV equation (\ref{kdv}) have been studied in literature. Thus we can use those known solutions to select out the $P_sT_d$ invariant solutions for the AB-KdV system (\ref{abkdv}). Here we list three special significant examples.
\subsection{$P_s T_d$-invariant multi-soliton solutions of the AB-KdV system}
For the KdV equation (\ref{kdv}), it is well known multiple soliton solution possesses the form \cite{Hirota}
\begin{eqnarray}
u=\frac{1}{2}(\ln F)_{xx},\label{solf}
\end{eqnarray}
\begin{eqnarray}
F=\sum_{\nu}\textrm{exp}\left(\sum_{j=1}^N\nu_j \xi_j+\sum_{1\leq j \leq l}^N\nu_j \nu_l \theta _{jl}\right), \label{f}
\end{eqnarray}
where the summation of $\nu$ should be done for all permutations of $\nu_i=0, 1, i=1, 2, \cdots, N$ and
\begin{eqnarray}
\xi_j=k_j x- k_j^3 t+\xi_{0j}, \qquad \textrm{exp}(\theta_{jl})=\left(\frac{k_j-k_l}{k_j+k_l}\right)^2. \label{xij}
\end{eqnarray}
It is clear that the solution (\ref{solf}) with (\ref{f}) and (\ref{xij}) is not $P_s T_d$ group invariant. To discuss the $N$-soliton solution to the AB-KdV system (\ref{abkdv}), we rewrite (\ref{xij}) as
\begin{eqnarray}
\xi_j =\eta_j-\frac{1}{2} \sum_{i=1}^{j-1}\theta_{ij}-\frac{1}{2}\sum_{i=j+1}^{N}\theta_{ji},
\end{eqnarray}
where
\begin{eqnarray}
\eta_j \equiv k_j\left(x-\frac{x_0}{2}\right)-k_j^3\left(t-\frac{t_0}{2}\right) +\eta_{0j},\label{etaj}
\end{eqnarray}
then the $N$-soliton solution of the KdV equation (\ref{kdv}) is
\begin{eqnarray}
u=\frac{1}{2}\left[\ln \sum_{\nu=-1, 1} K_{\nu} \cosh \left(\frac{1}{2}\sum_{j=1}^N \nu_j \eta_j \right) \right]_{xx},\label{pstdu}
\end{eqnarray}
where the summation of $\nu=\{\nu_1,\ \nu_2, \ \cdots, \ \nu_N\}$
should be done for all non-dual permutations of $i=1,\ 2, \ \cdots, \ N$ and
\begin{eqnarray}
K_{\nu}=\prod_{i>j}(k_i-\nu_i \nu_j k_j).\nonumber
\end{eqnarray}

From the expression (\ref{pstdu}), it is easy to see that
\begin{eqnarray}
A=u_{\eta_{0j}=0},\label{nsoliton}
\end{eqnarray}
solves the AB-KdV system (\ref{abkdv}). The solution (\ref{nsoliton}) with (\ref{etaj}) and (\ref{pstdu}) to the AB-KdV system (\ref{abkdv}) is $P_s T_d$ invariant because $A=u_{\eta_{0j}=0} =\hat{P_s}\hat{T_d} u_{\eta_{0j}=0}$. Then the multiple soliton solution $A=u_{\eta_{0j}=0}$ presents the $P_s T_d$ group invariant solution of the AB-KdV system (\ref{abkdv}).
\subsection{$P_s T_d$-invariant symmetry reduction solution to the AB-KdV system}
The knowledge of the symmetries is very useful to enhance our understanding of complex physical phenomena, to simplify and even completely solve the complicated problems. Furthermore, the study of symmetries has been manifested as one of the most important and powerful methods in almost every branch of science especially in physics and mathematics. It is particularly fundamental to find the symmetries of a nonlinear equation in the development of the theory of the integrable systems because of the existence of infinitely many symmetries \cite{lie} - \cite{bluman}.

It is easy to check that the common KdV equation (\ref{kdv}) possesses a special symmetry reduction
\begin{eqnarray}
u=t^{-\frac{2}{3}}U(\xi)^2+\frac{x}{24t}, \qquad \xi=t^{-\frac{1}{3}}x,\label{kdvre1}
\end{eqnarray}
where $U=U(\xi)$ satisfies
\begin{eqnarray}
U_{\xi \xi}+4 U^3+\frac{1}{6}U \xi+\frac{\alpha}{U^3}=0,\label{kdvre2}
\end{eqnarray}
which is equivalent to the Painlev\'{e} II equation.

In order to search the symmetry reduction solution for the AB-KdV system (\ref{abkdv}), we have to rewrite the function $U$ and $\xi$ in (\ref{kdvre1}) of the common KdV equation (\ref{kdv}) to satisfy the $P_sT_d$ invariant condition, so that the symmetry reduction solution of the AB-KdV system (\ref{abkdv}) is
\begin{eqnarray}
 A=\left(t-\frac{t_0}{2}\right)^{-\frac{2}{3}}U_1(\xi_1)^2+\frac{1}{24}\frac{2x-x_0}{2t-t_0},\quad \xi_1=\left(x-\frac{x_0}{2}\right) \left(t-\frac{t_0}{2}\right)^{-\frac{1}{3}},\label{piireduc1}
\end{eqnarray}
with $U_1(\xi_1)$ being a solution of
\begin{eqnarray}
U_{1\xi_1 \xi_1}+4 U_1^3+\frac{1}{6}U_1 \xi_1+\frac{\alpha}{U_1^3}=0.\label{piiredc2}
\end{eqnarray}

\subsection{$P_s T_d$-invariant soliton-cnoidal wave interaction solution of the AB-KdV system}
Many literatures \cite{louriccati} have studied the soliton-cnoidal wave interaction solutions for the nonlinear systems, such as the KdV equation, the KP equation etc. For example, a special soliton-cnoidal periodic wave interaction solution of the KdV equation (\ref{kdv}) reads as
\begin{eqnarray}
u=&& -\frac{1}{2}w_x^2\tanh^2(w)+\frac{1}{2}w_{xx}\tanh(w) +\frac{1}{3}w_x^2-\frac{1}{6}\frac{w_{xxx}}{w_x} -\frac{1}{24}\frac{w_t} {w_x}\nonumber \\ && \qquad +\frac{1}{8}\frac{w_{xx}^2}{w_x^2},\label{kdvsol1}
\end{eqnarray}
where $w$ is
\begin{eqnarray}
w=\frac{1}{2}k(x-c_1t)+\frac{1}{2}\textrm{arctanh}(m \textrm{sn}(k( x-c_2t),m)),\label{kdvsol2}
\end{eqnarray}
\begin{eqnarray}
k=\sqrt{\frac{c_1-c_2}{2(1-m^2)}},\label{kdvsol3}
\end{eqnarray}
with arbitrary constants $c_1$, $c_2$ and $m$.

It is clear that the solution (\ref{kdvsol1}) with (\ref{kdvsol2}) and (\ref{kdvsol3}) is $P_s T_d$ invariant. So the corresponding $P_s T_d$ invariant soliton-cnoidal periodic wave interaction solution of the AB-KdV system (\ref{abkdv}) is
\begin{eqnarray}
A=&& -\frac{1}{2}w_{1x}^2\tanh^2(w_1)+\frac{1}{2}w_{1xx}\tanh(w_1) +\frac{1}{3}w_{1x}^2 -\frac{1}{6}\frac{w_{1xxx}}{w_{1x}} -\frac{1}{24}\frac{w_{1t}} {w_{1x}}\nonumber \\ && \qquad +\frac{1}{8}\frac{w_{1xx}^2}{w_{1x}^2},\label{a1}
\end{eqnarray}
where $w_1$ is related to $\xi_1$ and $\xi_2$
\begin{eqnarray}
&& w_1=\frac{1}{2}k \xi_1+\frac{1}{2}\textrm{arctanh}(m \textrm{sn}(k \xi_2,m)),\qquad  k=\sqrt{\frac{c_1-c_2}{2(1-m^2)}},
\end{eqnarray}
\begin{eqnarray}
&& \xi_1=\left(x-\frac{x_0}{2}\right)-c_1\left(t-\frac{t_0}{2}\right), \qquad \xi_2=\left(x-\frac{x_0}{2}\right)-c_2\left(t-\frac{t_0}{2}\right),\label{a3}
\end{eqnarray}
with arbitrary constants $c_1$, $c_2$ and $m$.

\begin{figure}
\begin{center}
\subfigure{
\includegraphics[width=0.35\textwidth]{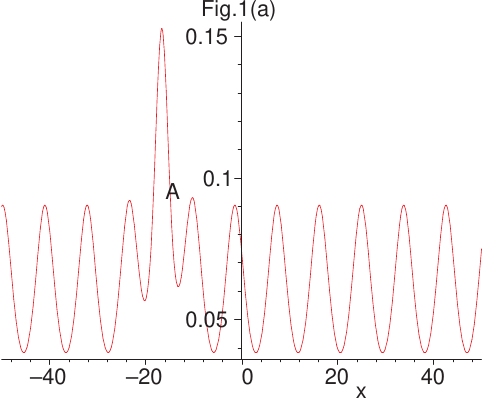}
\includegraphics[width=0.35\textwidth]{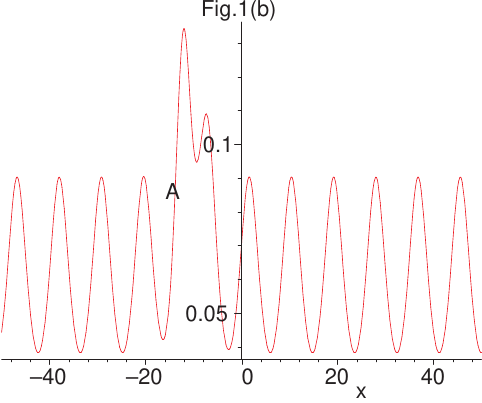}}
\subfigure{
\includegraphics[width=0.35\textwidth]{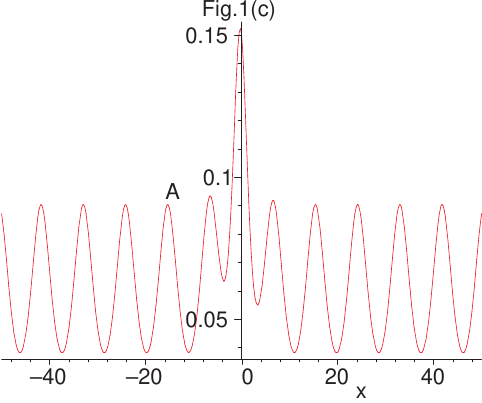}
\includegraphics[width=0.35\textwidth]{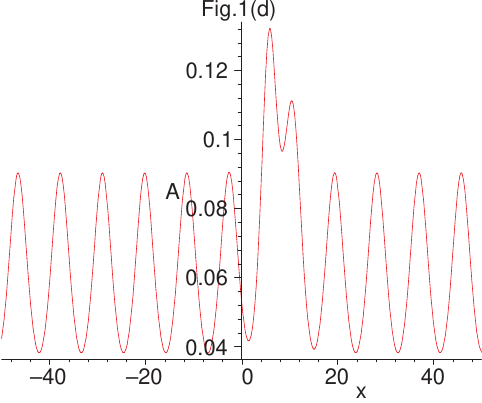}}
\caption{The exhibition of the $P_sT_d$ invariant soliton-cnoidal wave interaction solution of eqs. (\ref{a1}) - (\ref{a3}) with parameters selections eq. (\ref{a4}) at times (a) $t=-8$, (b) $t=-5$, (c) $t=0$ and (d) $t=4$, respectively.\label{fig1}}
\end{center}
\end{figure}

Figure \ref{fig1} shows a special $P_sT_d$ invariant soliton-cnoidal wave interaction solution eqs. (\ref{a1}) - (\ref{a3}) of the AB-KdV system (\ref{abkdv}) with parameters selected as
\begin{eqnarray}
c_1=2, \qquad c_2=1, \qquad m=0.2, \qquad x_0=1, \qquad t_0=1.\label{a4}
\end{eqnarray}

Many $P_s T_d$-invariant solutions of the AB-KdV system (\ref{abkdv}) have been found out with the help of the common KdV equation (\ref{kdv}) by using the $P_sT_d$ principle, including multi-soliton solution, symmetry reduction solution and soliton-cnoidal wave interaction solution. The $P_sT_d$ invariant solutions mean that the event happened at $\{x,\ t\}$ will happen also at $\{x',\ t'\} \equiv \{-x+x_0,\ -t+t_0\}$. Then new problems arise up: Are there $P_s T_d$ symmetry breaking solutions of the AB-KdV system (\ref{abkdv})? How to obtain $P_s T_d$ symmetry breaking solutions of the AB-KdV system (\ref{abkdv})? It is important and interesting but difficult to solve the problems.
\section{$P_s T_d$ Symmetry Breaking Solutions of the AB-KdV System}
In this section, we aim to search for the $P_s T_d$ symmetry breaking solutions of the AB-KdV system (\ref{abkdv}).

Motivated by the $P_s T_d$ symmetry solutions of the AB-KdV system (\ref{abkdv}) being obtained from those of the original KdV equation (\ref{kdv}) using $P_sT_d$ principle, we now construct the $P_s T_d$ symmetry breaking solutions of the AB-KdV system (\ref{abkdv}) from a coupled KdV system (cKdV system)
\begin{eqnarray}
&& u_t+u_{xxx}+3(u+v)(v_x+3u_x)=0, \label{ckdv1}
\\ && v_t+v_{xxx}+3(u+v)(u_x+3v_x)=0, \label{ckdv2}
\end{eqnarray}
with $u=u(x,t)$ and $v=v(x,t)$ being a normal coupled KdV system. It can be directly found that the cKdV system (\ref{ckdv1}) - (\ref{ckdv2}) possesses a special reduction of $v=u(-x,-t)$ which makes the coupled KdV system (\ref{ckdv1})  - (\ref{ckdv2}) being reduced to one equation
\begin{eqnarray}
&& u_t+u_{xxx}+3(u+v)(v_x +3u_x)=0,\label{redckdv}
\end{eqnarray}
with $v=u(-x,-t)$. The single equation (\ref{redckdv}) with the constraint condition $v=u(-x,-t)$ can be considered as a special case of the AB-KdV system (\ref{abkdv}) with $x_0=t_0=0$. That implies to look for the exact solutions of the AB-KdV system (\ref{abkdv}) is equivalent to find out the exact solutions of the coupled KdV equation (\ref{ckdv1}) - (\ref{ckdv2}). Once the exact solutions of the cKdV system (\ref{ckdv1}) - (\ref{ckdv2}) are known, by using the special reduction of $v=u(-x,-t)$ to these solutions, the exact solutions of the AB-KdV system (\ref{abkdv}) may be presented. Thus it is possible to find out the $P_s T_d$ symmetric breaking solutions of the AB-KdV system (\ref{abkdv}).

In this section, we will show details on how to obtain the $P_s T_d$ symmetric breaking solutions of the AB-KdV system (\ref{abkdv}) with the help of the coupled KdV system (\ref{ckdv1}) - (\ref{ckdv2}), including single soliton solutions, singular soliton solutions and cnoidal wave solutions with the help of the coupled KdV system (\ref{ckdv1})- (\ref{ckdv2}).
\subsection{single soliton solutions of the AB-KdV system}
As we have known, one of the notable features of the integrable systems is the possessing of soliton solutions. Many reliable methods are used in the literature to examine the soliton solutions of the integrable nonlinear evolution equations. The Hirota bilinear method \cite{hirota}, the Backlund transformation method, the inverse scattering method \cite{ist}, the Painlev\'{e} analysis, and others are effectively used to determine soliton solutions for completely integrable equations. Among all the methods, the $\tanh$ function expansion approach \cite{lanjpa, faneg, new1} is one of the simple and effective methods to search for the soliton solutions for the integrable equations.

To find the single soliton solutions of the cKdV system (\ref{ckdv1})-(\ref{ckdv2}), by means of the $\tanh$ function expansion approach, it is direct to assume the single soliton solution of the cKdV system (\ref{ckdv1}) - (\ref{ckdv2}) has the form
\begin{eqnarray}
&& u=a_0+a_1 \tanh(k x+ct) +a_2 \tanh^2(k x+c t), \nonumber \\ && v=b_0+b_1 \tanh(k x+ct) +b_2 \tanh^2(k x+c t),\label{ckdvsol1}
\end{eqnarray}
by balancing the powers of $\tanh$ function of the highest order derivative term and nonlinear term in cKdV system, with the parameters $a_i$ and $b_i$, $i=0,\ 1,\ 2$, $k$, $c$ being real arbitrary constants to be determined.

Substituting the ansatz (\ref{ckdvsol1}) into the cKdV system (\ref{ckdv1}) - (\ref{ckdv2}) and vanishing all the coefficients of $\tanh^i(k x+c t)$ for $i$, then determining the real constants for $a_i$ and $b_i$, $i=0,\ 1,\ 2$, $k$, $c$, the single soliton solution of the the cKdV system (\ref{ckdv1})-(\ref{ckdv2}) reads as
\begin{eqnarray}
&& u=a_1 \tanh(k x-4k^3 t)-\frac{1}{2} k^2 \tanh^2(k x-4k^3 t) +k^2-b_0, \nonumber \\ && v=-a_1 \tanh(k x-4k^3 t)-\frac{1}{2} k^2 \tanh^2(k x-4k^3 t) +b_0,\label{ckdvs1}
\end{eqnarray}
where $k$, $a_1$ and $b_0$ are arbitrary constants.

To explore the exact solutions of the AB-KdV system (\ref{ckdv1}) - (\ref{ckdv2}), we put the special reduction $v=u(-x, -t)$ on the special single soliton solution (\ref{ckdvs1}) which leads to a constraint condition
\begin{eqnarray}
b_0 = k^2-b_0.\label{ccbk}
\end{eqnarray}
The constraint condition means $b_0$ has to be fixed as
\begin{eqnarray}
b_0=\frac{k^2}{2}, \nonumber
\end{eqnarray}
which means the special case of the AB-KdV system (\ref{redckdv}) possesses the single soliton solution
\begin{eqnarray}
&&u= \frac{k^2}{2}+ a_1 \tanh (k x-4 k^3 t) -\frac{1}{2} k^2 \tanh^2 (k x-4 k^3 t),\nonumber \\ \label{ssabkdv} &&v=u(-x,\ -t).\label{sss1}
\end{eqnarray}

The coupled KdV system (\ref{redckdv}) can be considered as a special case of the AB-KdV system (\ref{abkdv}) with $x_0=0$, $t_0=0$, so the corresponding single soliton solution of the AB-KdV system (\ref{abkdv}) is directly obtained by rewriting some parameters to satisfy the property of $P_sT_d$ invariant
\begin{eqnarray}
A= &&\frac{k^2}{2}+ a_1 \tanh \left[\left(x-\frac{x_0}{2}\right)-4 k^3 \left(t-\frac{t_0}{2}\right)\right] -\frac{1}{2} k^2 \tanh^2 \left[k \left(x-\frac{x_0}{2}\right)\nonumber \right. \\ &&\qquad -4 k^3 \left. \left(t-\frac{t_0}{2}\right)\right], \label{ssabkdv} \nonumber \\ B=&&A(-x+x_0,\ -t+t_0),\label{sss1}
\end{eqnarray}
with $a_1$ and $k$ being two arbitrary constants.

It should be pointed out that the single soliton solution (\ref{ssabkdv}) for $a_1 \neq 0$ is $P_sT_d$ symmetric breaking because by applying the operator $\hat{P_s} \hat{T_d}$ on the solution (\ref{sss1}), we find $\hat{P_s} \hat{T_d} A \neq A$. If $a_1=0$, the solution (\ref{ssabkdv}) becomes a $P_sT_d$ invariant solution.
\subsection{infinitely many singular soliton solutions}
Now the $P_sT_d$ symmetric breaking single soliton solution of the AB-KdV has been given with the help of the cKdV system (\ref{ckdv1}) - (\ref{ckdv2}). Furthermore, we continue on searching for the infinitely many soliton solutions of the AB-KdV system (\ref{abkdv}). We still start from the cKdV system (\ref{ckdv1}) - (\ref{ckdv2}).

To construct infinitely many soliton solutions for the cKdV system (\ref{ckdv1}) - (\ref{ckdv2}), we introduce two undetermined functions $U(\eta,\ t)$ and $V(\eta,\ t)$ where $\eta \equiv \tanh(k x-4k^3 t)$ with $k$ being arbitrary constant. Based on the known results, we assume the infinitely many soliton solutions for the cKdV system (\ref{ckdv1}) - (\ref{ckdv2}) are
\begin{eqnarray}
&& u=a_1 U(\eta, t)-\frac{1}{2} k^2 \eta^2+k^2-b_0, \nonumber \\ && v=-a_1 V(\eta, t)-\frac{1}{2} k^2 \eta^2+b_0, \label{mss} \label{ckdvsol2}
\end{eqnarray}
where $a_1$ and $b_0$ are arbitrary constants to be determined.

Substituting (\ref{ckdvsol2}) into the cKdV system (\ref{ckdv1}) - (\ref{ckdv2}), we obtain the determining equations for $U(\eta,\ t)$ and $V(\eta,\ t)$
\begin{eqnarray}
U_t=&& 3 k(\eta^2-1) [3 U+3 V+k^2 (\eta^2-1)] U_{\eta}+3 k (\eta^2-1) [U+V \nonumber \\ \qquad &&+k^2 (\eta^2-1)]V_{\eta} + 6 \eta k^3(\eta^2-1)^2 U_{\eta \eta}+k^3 (\eta^2-1)^3 U_{\eta \eta \eta} \nonumber \\ \qquad &&-12 \eta k^3 (\eta^2-1)(U+V), \nonumber \\V_t=&&3 k(\eta^2-1) [3 U+3 V+k^2 (\eta^2-1)] V_{\eta}+3 k (\eta^2-1) [U+V\nonumber \\ \qquad &&+k^2 (\eta^2-1)]U_{\eta}+ 6 \eta k^3(\eta^2-1)^2 V_{\eta \eta}+k^3 (\eta^2-1)^3 V_{\eta \eta \eta} \nonumber \\ \qquad &&-12 \eta k^3 (\eta^2-1)(U+V).\label{uveta}
\end{eqnarray}
The equation (\ref{uveta}) for $U(\eta, t)$ and $V(\eta, t)$ is complicated nonlinear coupled system which is difficult to solve. In order to seek for the exact solutions of the complicated system (\ref{uveta}), we take a special case of $U=-V$. It is clearly seen that the coupled system (\ref{uveta}) is reduced to one linear equation
\begin{eqnarray}
V_t=6 k^3 \eta (\eta^2-1)^2 V_{\eta \eta}+k^3 (\eta^2-1)^3 V_{\eta \eta \eta}, \label{vequation}
\end{eqnarray}
which can be solved by means of the variable separation approach. It is not difficult to verify that eq. (\ref{vequation}) possesses a special solution
\begin{eqnarray}
V=&& \sum_{i=1}^N a_i \left\{\left(\frac{1-\eta}{1+\eta}\right)^{c_i}(\eta+c_i)\textrm{exp}\left[8c_i k^3(c_i^2-1) t\right]\right.\nonumber \\ &&\qquad +\left. \left(\frac{1-\eta}{1+\eta}\right)^{-c_i}(\eta-c_i)\textrm{exp}[-8c_i k^3(c_i^2-1) t] \right\} +c_{01}\eta,\label{solv}
\end{eqnarray}
with $a_i$, $c_i$ and $c_{01}$ being arbitrary constants which is analytical only for $c_i=0$, $i=1, \cdots, N$.

Thus the corresponding infinitely many singular soliton solution of the cKdV system (\ref{ckdv1}) - (\ref{ckdv2}) is obtained by substituting the known results into (\ref{mss})
\begin{eqnarray}
&& u=-a_1 V(\eta, t)-\frac{1}{2} k^2 \eta^2+k^2-b_0, \nonumber \\ && v=-a_1 V(\eta, t)-\frac{1}{2} k^2 \eta^2+b_0,\label{imsss}
\end{eqnarray}
where $V(\eta, t)$ is shown in (\ref{solv}) and $\eta \equiv \tanh(k x-4k^3 t)$.

To present the exact solution of the AB-KdV system (\ref{abkdv}), after putting the constraint condition $v=u(-x,\ -t)$ in (\ref{imsss}) by some calculations, it is easy to find out the corresponding infinitely many singular soliton solution of the AB-KdV system (\ref{abkdv}) is derived
\begin{eqnarray}
A=&&c_{01}\eta_1-\frac{1}{2}k^2 \eta_1^2 +\frac{1}{2}k^2+ \sum_{i=1}^N a_i \left[\left(\frac{1-\eta_1}{1+\eta_1}\right)^{c_i}(\eta_1+c_i)\textrm{exp}[8c_i k^3(c_i^2-1) t]\right.\nonumber \\ &&\qquad +\left. \left(\frac{1-\eta_1}{1+\eta_1}\right)^{-c_i}(\eta_1-c_i)\textrm{exp}[-8c_i k^3(c_i^2-1) t] \right] \nonumber \\
\eta_1=&& \tanh\left[ k \left(x-\frac{x_0}{2}\right)-4 k^3 \left(t-\frac{t_0}{2}\right)\right],\label{abkdvsol1}
\end{eqnarray}
with $a_i$, $c_i$, $c_{01}$ and $k$ being arbitrary constants. The solution (\ref{abkdvsol1}) is analytical only for $c_i=0$, $i=1, \cdots, N$.

It is interesting that the infinitely many singular soliton solution (\ref{abkdvsol1}) is also anti-$P_s T_d$ symmetric for $c_{01} \neq 0$.
\subsection{cnoidal wave solutions}
Except for the soliton solutions, we also explore the cnoidal wave solutions of the AB-KdV system (\ref{abkdv}). As stated before, we first start from the cKdV system (\ref{ckdv1}) - (\ref{ckdv2}) to construct the cnoidal wave solutions, and then apply the constraint reduction $v=u(-x, -t)$ to these solutions of cKdV system (\ref{ckdv1}) - (\ref{ckdv2}) to find the solutions of the AB-KdV system (\ref{abkdv}).

Jacobi elliptic function expansion method is one of the most common approaches to explore the cnoidal wave solutions of a nonlinear system. Using the Jacobi elliptic function expansion method, we assume the cKdV system (\ref{ckdv1}) - (\ref{ckdv2}) possesses the cnoidal wave solution \cite{lounijmp}
\begin{eqnarray}
&& u=a_0 +a_1\textrm{sn}(k x + ct,m)+a_2  \textrm{sn}^2(k x + ct,m), \nonumber \\ && v=b_0 +b_1\textrm{sn}(k x + ct,m)+b_2  \textrm{sn}^2(k x + ct,m), \label{ckdvsol3}
\end{eqnarray}
where $a_0$, $a_1$ $b_0$, $b_1$, $k$ and $c$ are undetermined real constants. Substituting (\ref{ckdvsol3}) into the cKdV system (\ref{ckdv1})-(\ref{ckdv2}) and vanishing all the coefficients of $\textrm{sn}^i(k x + ct,m)$ for different $i$, the exact solutions of (\ref{ckdv1}) - (\ref{ckdv2}) can be read as
\begin{eqnarray}
&& u=a_1\textrm{sn}[k x -2k^3(m^2+1)t,m] -\frac{1}{2} k^2  \textrm{sn}^2[k x -2k^3(m^2+1)t,m]+a_0, \nonumber \\ && v=-a_1 \textrm{sn}[k x -2k^3(m^2+1)t,m] -\frac{1}{2} k^2 \textrm{sn}^2[k x -2k^3(m^2+1)t,m]\nonumber \\ && \qquad +\frac{1}{2} k^2 m^2+\frac{1}{2} k^2-a_0, \label{ckdvcw}
\end{eqnarray}
with $a_0$ and $a_1$ being arbitrary constant.

To obtain the exact solutions of the AB-KdV system (\ref{abkdv}), we apply the constraint $v=u(-x,-t)$ to the solutions (\ref{ckdvcw}) of the cKdV system (\ref{ckdv1}) - (\ref{ckdv2}) which leads to
\begin{eqnarray}
a_0=\frac{1}{2} k^2 m^2+\frac{1}{2} k^2-a_0.
\end{eqnarray}
That means $a_0$ should be fixed as
\begin{eqnarray}
a_0=\frac{1}{4} k^2 m^2+\frac{1}{4} k^2.
\end{eqnarray}

We have known that $v=u(-x,-t)$ in the cKdV system (\ref{ckdv1})-(\ref{ckdv2}) leads to a special case of the AB-KdV system (\ref{abkdv}) with $x_0=t_0=0$, thus after rewriting the parameters, the corresponding cnoidal wave  solution of AB-KdV system (\ref{abkdv}) is found out
\begin{eqnarray}
&& A=a_1\zeta-\frac{1}{2} k^2  \zeta^2+\frac{1}{4} k^2 m^2+\frac{1}{4} k^2, \nonumber \\ && \zeta=\textrm{sn}\left[k \left(x-\frac{x_0}{2}\right) -2k^3(m^2+1)\left(t-\frac{t_0}{2}\right),m \right],\label{abkdvsol2}
\end{eqnarray}
with $a_1$, $k$ and $m$ being arbitrary constants. It is obvious the cnoidal wave solution (\ref{abkdvsol2}) of the AB-KdV system (\ref{abkdv}) is also anti-$P_s T_d$ symmetric for $a_1 \neq 0$.

Moreover, it is known that when the modulus $m$ of the Jacobi elliptic function tends to $1$, the conoidal wave tends to a soliton or solitary wave solution. It can be easily verified that if $m=1$, the cnoidal wave solution (\ref{abkdvsol2}) will become the single soliton solution (\ref{sss1}).
\subsection{infinitely many periodic wave solutions}
We next consider the infinitely many periodic wave solutions of the cKdV system (\ref{ckdv1}) - (\ref{ckdv2}) and the AB-KdV system (\ref{abkdv}). In order to obtain the infinitely many periodic wave solutions of the cKdV system (\ref{ckdv1}) - (\ref{ckdv2}), we substitute the ansatz
\begin{eqnarray}
&& u=F(\theta,t)-\frac{1}{2}k^2 m^2 \theta^2+a_0,\nonumber \\ && v=G(\theta,t)-\frac{1}{2} k^2 m^2 \theta^2+\frac{1}{2} k^2 m^2+\frac{1}{2} k^2-a_0,
\end{eqnarray}
into the cKdV system (\ref{ckdv1}) - (\ref{ckdv2}) to determine the functions of $F(\theta, t)$ and $G(\theta, t)$, where $\theta \equiv \textrm{sn}[k x -2k^3(m^2+1)t,m]$ with $k$ being arbitrary constant. It is not difficult to find the determining equations of $F(\theta,t)$ and $G(\theta,t)$ are
\begin{eqnarray}
&&F_t= -\frac{3}{2}z_0k(6 F+6G+ k^2 z_1)F_{\theta}-\frac{3}{2}z_0 k(2 F+2G+ k^2 z_1)G_{\theta}\nonumber \\ && \qquad -3z_0 z_1 k^3 \theta F_{\theta \theta}-k^3 z_0^3F_{\theta \theta \theta} +12 z_0 \theta m^2 k^3 (F+G), \nonumber \\&&G_t= -\frac{3}{2}z_0k(6 F+6G+ k^2 z_1)G_{\theta}-\frac{3}{2}z_0 k(2 F+2G+ k^2 z_1)F_{\theta}\nonumber \\ && \qquad-3z_0 z_1 k^3 \theta G_{\theta \theta}-k^3 z_0^3 G_{\theta \theta \theta} +12 z_0 \theta m^2 k^3 (F+G),\label{FG}
\end{eqnarray}
with
\begin{eqnarray}
&&z_0 =\sqrt{1-\theta^2}\sqrt{1-m^2\theta^2},\qquad z_1=m^2-2 m^2 \theta^2+1.\label{z0z1}
\end{eqnarray}

Once eqs. (\ref{FG}) and (\ref{z0z1}) are solved, the infinitely many periodic wave solution of the cKdV system (\ref{ckdv1}) - (\ref{ckdv2}) is obtained. But it is difficult to find the exact solutions for such a complicated coupled nonlinear system. Fortunately, if we introduce a function $H(\theta, t)$ satisfying $H(\theta,t)=F(\theta, t)+G(\theta, t)$, the determining coupled system (\ref{FG}) becomes
\begin{eqnarray}
&&H_t=24 z_0 \theta m^2 k^3 H-3z_0k(4 H+ k^2 z_1)H_{\theta} -3z_0 z_1 k^3 \theta H_{\theta \theta}-k^3 z_0^3 H_{\theta \theta \theta}, \label{H} \\
&&G_t= 12 z_0 \theta m^2 k^3 H-\frac{3}{2}z_0k(2H+z_1)H_{\theta}-6k z_0 H G_{\theta} -3 z_0 z_1 \theta k^3 G_{\theta \theta}\nonumber \\ && \qquad -k^3 z_0^3 G_{\theta \theta \theta}. \label{G}
\end{eqnarray}

It is still not easy to solve the coupled system of (\ref{H}) and (\ref{G}), but it can be directly seen that the equation (\ref{H}) possesses a special solution of $H(\theta,t)=0$. And then (\ref{G}) is reduced to a linear equation reading as
\begin{eqnarray}
G_t=-3 z_0 z_1 \theta k^3 G_{\theta \theta}-k^3 z_0^3 G_{\theta \theta \theta}, \label{G0}
\end{eqnarray}
which can be solved by means of the variable separation approach. Here we don't show the concrete results in detail.

According to the known results, we can directly write down the infinitely many periodic wave solutions of the cKdV system (\ref{ckdv1}) - (\ref{ckdv2}) in two cases:\\
\textbf{Case 1.} For $H(\theta, t)\neq 0$, the solution of the cKdV system (\ref{ckdv1}) - (\ref{ckdv2}) is
\begin{eqnarray}
&& u=-G(\theta, t)+H(\theta, t)-\frac{1}{2}k^2 m^2 \theta^2+a_0,\nonumber \\ && v=G(\theta,t)-\frac{1}{2} k^2 m^2 \theta^2+\frac{1}{2} k^2 m^2+\frac{1}{2} k^2-a_0,
\end{eqnarray}
where $H(\theta, t)$ and $G(\theta, t)$ are constrainted by the coupled system (\ref{H}) and (\ref{G}), with $z_0$ and $z_1$ satisfying (\ref{z0z1}), $a_0$ and $k$ being arbitrary constants. \\
\textbf{Case 2.} A special solution of the cKdV system (\ref{ckdv1}) - (\ref{ckdv2}) for $H(\theta, t)=0$ is
\begin{eqnarray}
&& u=-G(\theta, t)-\frac{1}{2}k^2 m^2 \theta^2+a_0,\nonumber \\ && v=G(\theta,t)-\frac{1}{2} k^2 m^2 \theta^2+\frac{1}{2} k^2 m^2+\frac{1}{2} k^2-a_0,
\end{eqnarray}
where $G(\theta, t)$ is given by (\ref{G0}), with $z_0$ and $z_1$ satisfying (\ref{z0z1}). Once the solution of (\ref{G0}) is presented by means of the variable separation approach, it is possible to find more soliton solutions and periodic wave solutions of the cKdV system (\ref{ckdv1}) - (\ref{ckdv2}).

As the exact solutions of the cKdV system (\ref{ckdv1}) - (\ref{ckdv2}) have been obtained, we can rewrite the solutions to satisfy the AB-KdV system. Here we just write down the results.

The AB-KdV system possesses the infinitely many periodic wave solutions as follow:\\
\textbf{Case 1.} If $\{G_1(\theta_1,t),\ H_1(\theta_1,t)\}$ with $H_1(\theta, t)\neq 0$ satisfies the constraint equations
\begin{eqnarray}
H_{1t}&&=-3z_0k(4 H_1+ k^2 z_1)H_{1\theta_1} -3z_0 z_1 k^3 \theta H_{1 \theta_1 \theta_1}-k^3 z_0^3 H_{1 \theta_1 \theta_1 \theta_1}\nonumber \\ && \qquad+24 z_0 \theta_1 m^2 k^3 H_1, \label{H1}
 \nonumber \\
G_{1t}&&= -\frac{3}{2}z_0k(2H_1+z_1)H_{1 \theta_1}-6k z_0 H_1 G_{1 \theta_1}-3 z_0 z_1 \theta_1 k^3 G_{\theta_1 \theta_1}\nonumber \\ && \qquad -k^3 z_0^3 G_{1 \theta_1 \theta_1 \theta_1}+12 z_0 \theta_1 m^2 k^3 H_1. \label{G1}
\end{eqnarray}
then
\begin{eqnarray}
&& A=-G_1(\theta_1,t)+H_1(\theta_1,t)-\frac{1}{2}k^2 m^2 \theta_1^2+\frac{1}{4}k^2 m^2+\frac{1}{4}k^2,\nonumber \\ && B= G_1(\theta_1,t)-\frac{1}{2}k^2 m^2 \theta_1+\frac{1}{4} k^2 m^2+\frac{1}{4} k^2,\nonumber \\ && \theta_1=\textrm{sn}\left[k \left(x-\frac{x_0}{2}\right) -2k^3(m^2+1)\left(t-\frac{t_0}{2}\right),m \right],\label{sol4}
\end{eqnarray}
is an infinitely many periodic wave solution of the AB-KdV system (\ref{abkdv}).\\
\textbf{Case 2.} A special solution of the AB-KdV system (\ref{abkdv}) is
\begin{eqnarray}
A&&=-G_1(\theta_1,t)-\frac{1}{2}k^2 m^2 \theta_1^2+\frac{1}{4}k^2 m^2+\frac{1}{4}k^2,\nonumber \\ B&&= G_1(\theta_1,t)-\frac{1}{2}k^2 m^2 \theta_1^2+\frac{1}{4} k^2 m^2+\frac{1}{4} k^2,\nonumber \\  \theta_1&& =\textrm{sn}\left[k \left(x-\frac{x_0}{2}\right) -2k^3(m^2+1)\left(t-\frac{t_0}{2}\right),m \right],\label{abkdvsol3}
\end{eqnarray}
where $G_1(\theta_1,t)$ is the solution of
\begin{eqnarray}
G_{1t}=-3z_0 z_1 \theta_1 k^3 G_{1\theta_1 \theta_1}-k^3 z_0^3 G_{1\theta_1 \theta_1 \theta_1}, \label{G01}
\end{eqnarray}
with $z_0$ and $z_1$ satisfying (\ref{z0z1}).

It should be pointed out that all the anti-symmetric ($H_1(-\theta_1, -t)=-H_1(\theta_1, t)$, $G_1(-\theta_1, -t)=-G_1(\theta_1, t)$) solutions of $H_1$ and $G_1$ will lead to the anti-$P_sT_d$ symmetric solutions of the AB-KdV system (\ref{abkdv}).

In this section, we successfully construct the $P_sT_d$ symmetric breaking solutions of the AB-KdV system (\ref{abkdv}), including single soliton solutions, infinitely many singular soliton solutions, cnoidal wave solutions and infinitely many periodic wave solutions with the help of a couple KdV system. The AB-KdV system (\ref{abkdv}) possesses rich solution structures and it is possible to find out more soliton solutions and periodic wave solutions due to the rich solution structures of the cKdV system (\ref{ckdv1}) - (\ref{ckdv2}).
\section{Lie Point Symmetry and Symmetry Reduction}
The knowledge of the symmetries is very useful to enhance our understanding of complex physical phenomena, to simplify and even completely solve the complicated problems. Furthermore, the study of symmetries has been manifested as one of the most important and powerful methods in almost every branch of science especially in physics and mathematics. It is particularly fundamental to find the symmetries of a nonlinear equation in the development of the theory of the integrable systems because of the existence of infinitely many symmetries \cite{lie}- \cite{bluman}. Recent studies show that symmetry reduction approach with nonlocal symmetries and residual symmetries has been successfully used to find some types of interaction solutions of nonlinear excitations for a number of integrable system \cite{gaoloutang} - \cite{jincomm}. In this section, we will discuss the Lie point symmetry and symmetry reduction solutions of the AB-KdV system (\ref{abkdv}) with the help of the cKdV system (\ref{ckdv1}) - (\ref{ckdv2}).

A symmetry, $\sigma$ is defined by a solution of its linearized equation. For the cKdV system (\ref{ckdv1}) - (\ref{ckdv2}), its symmetries are solutions of
\begin{eqnarray}
&& \sigma^u_t+\sigma^u_{xxx}+ 3 (\sigma^u+\sigma^v)(v_x+3 u_x)+3(u +v)(\sigma^v_x+3 \sigma^u_x)=0, \nonumber \\ && \sigma^v_t+\sigma^v_{xxx}+ 3 (\sigma^u+\sigma^v)(u_x+3v_x)+3(u +v)(\sigma^u_x+3\sigma^v_x)=0. \label{susv}
\end{eqnarray}
That means the cKdV system (\ref{ckdv1})-(\ref{ckdv2}) is form-invariant under the transformation
\begin{eqnarray}
\left(\begin{array}{c}
  u \\
  v
\end{array}\right)\rightarrow
\left( \begin{array}{c}
u+\epsilon \sigma^u \\
v+\epsilon \sigma^v
\end{array}\right),
\end{eqnarray}
where $\epsilon$ is an infinitesimal parameter.

The generalized Lie point symmetries of the cKdV system (\ref{ckdv1})-(\ref{ckdv2}) possess the form
\begin{eqnarray}
\left(\begin{array}{c}
        \sigma^u \\
        \sigma^v
      \end{array}
\right)=&& X(x,t,u,v)\left(\begin{array}{c}
                    u_x \\
                    v_x
                  \end{array}\right)
+T(x,t,u,v)\left(\begin{array}{c}
             u_t \\
             v_t
           \end{array}\right) \nonumber \\ && \qquad
-\left(\begin{array}{c}
   U(x,t,u,v) \\
   V(x,t,u,v)
 \end{array}\right),\label{susv10}
\end{eqnarray}
where $X(x,t,u,v)$, $T(x,t,u,v)$, $U(x,t,u,v)$ and $V(x,t,u,v)$ are the functions of the indicated variables.

Substituting (\ref{susv10}) into the symmetry definition (\ref{susv}) and using the cKdV system (\ref{ckdv1}) - (\ref{ckdv2}) to eliminate nonindependent quantities, $u_t$ and $v_t$, we can get a set of determining equations for $X$, $T$, $U$ and $V$. Solving the determining equation system, the Lie point symmetry can be obtained. It is easy to verify that the cKdV system (\ref{ckdv1})-(\ref{ckdv2}) possesses the Lie point symmetries
\begin{eqnarray}
&& \sigma^u=(c_1 x+c_2)u_x+(3 c_1 t+c_3)u_t+(2 c_1+c_4)u -c_4v-c_0, \nonumber \\ && \sigma^v=(c_1 x+c_2)v_x+(3 c_1 t+c_3)u_t+(2 c_1+c_4)v -c_4u +c_0,
\end{eqnarray}
where $c_i$, $i=0, 1, 2,3,4$ are arbitrary constants.

To find symmetry reductions of a nonlinear system means to find the group invariant solutions which is guaranteed by $\sigma^u=\sigma^v=0$,
\begin{eqnarray}
&& 0=(c_1 x+c_2)u_x+(3 c_1 t+c_3)u_t+(2 c_1+c_4)u -c_4v-c_0, \nonumber \\ && 0=(c_1 x+c_2)v_x+(3 c_1 t+c_3)u_t+(2 c_1+c_4)v -c_4u +c_0.\label{uv0}
\end{eqnarray}
Solving the group invariant conditions (\ref{uv0}) and the cKdV system (\ref{ckdv1}) - (\ref{ckdv2}), we can get the symmetry reduction solutions. Here we just list a special reduction solution because we aim to find the symmetry reductions of the AB-KdV system (\ref{abkdv}).

If $\{U(\xi),\ V(\xi)\}$ is the solution of the reduction equation
\begin{eqnarray}
&& U_{\xi \xi \xi}=\left(\frac{1}{3}\xi-6V\right)U_{\xi}-d U, \qquad V_{\xi \xi \xi}=\left(\frac{1}{3}\xi-12V\right)V_{\xi}+\frac{2}{3}V,\label{reduction1}
\end{eqnarray}
then
\begin{eqnarray}
&& u=t^d U+\frac{1}{2}t^{-\frac{2}{3}}V+C_0, \qquad v=-t^d U+\frac{1}{2}t^{-\frac{2}{3}}V-C_0, \label{reduction2}
\end{eqnarray}
is a solution of the cKdV system (\ref{ckdv1}) - (\ref{ckdv2}), where $d$ and $C_0$ are arbitrary constants with the independent variable $\xi=x t^{-\frac{1}{3}}$.

Now we have obtained the symmetry reduction (\ref{reduction1}) and (\ref{reduction2}) of the cKdV system (\ref{ckdv1}) - (\ref{ckdv2}), then we will utilize the results of cKdV system (\ref{ckdv1}) - (\ref{ckdv2}) to the AB-KdV system (\ref{abkdv}). Using the constraint condition $v=u(-x,\ -t)$, then rewriting the parameters to keep the $P_s T_d$ invariant of the results to satisfy the AB-KdV system (\ref{abkdv}), the symmetry reduction solution of the AB-KdV system (\ref{abkdv}) can be stated in the following theorem.

\textbf{Theorem 1} (Symmetry reduction theorem of the AB-KdV system (\ref{abkdv})). If $\{U_1(\xi_1),\ V_1(\xi_1)\}$ is a solution of the reduction equation
\begin{eqnarray}
&& U_{1 \xi_1 \xi_1 \xi_1}=\left(\frac{1}{3}\xi_1 -6 V_1\right)U_{1\xi_1}-\frac{2n+1}{2m+1}U_1,\nonumber \\ && V_{1\xi_1 \xi_1 \xi_1}=\left(\frac{1}{3}\xi_1-12V_1\right)V_{1\xi_1}+\frac{2}{3}V_1,\label{abreduc1}
\end{eqnarray}
then
\begin{eqnarray}
&& A=\left(t-\frac{t_0}{2}\right)^\frac{2n+1}{2m+1}U_1(\xi_1)+\frac{1}{2}\left(t-\frac{t_0}{2}\right)^{-\frac{2}{3}} V_1(\xi_1), \nonumber \\ && \xi_1=\left( x-\frac{x_0}{2} \right) \left(t- \frac{t_0}{2} \right) ^{-\frac{1}{3}}, \label{abreduc2}
\end{eqnarray}
with $m$ and $n$ being arbitrary integers, is a reduction solution of the AB-KdV system (\ref{abkdv}).

We should point out that the reduction (\ref{abreduc1}) has rich and various solution structures. It can be clearly seen that for any given $V_1(\xi_1)$, $U_1(\xi_1)$ is linear for the reduction (\ref{abreduc1}) which is easy to solve. For example, if $V_1=0$, the reduction equation (\ref{abreduc1}) is reduced to
\begin{eqnarray}
U_{1 \xi_1 \xi_1 \xi_1}=\frac{1}{3}\xi_1 U_{1\xi_1}-\frac{2n+1}{2m+1}U_1, \label{abreduc11}
\end{eqnarray}
which has many types of solutions due to the different choice of $m$ and $n$. Here we list two examples.\\
\textbf{Example 1.} If we take $m=n=0$, then the solution of (\ref{abreduc11}) is
\begin{eqnarray}
U_1=C_1 (\xi_1^3-6), \quad \xi_1=\left(x-\frac{x_0}{2}\right)\left(t-\frac{t_0}{2}\right)^{-\frac{1}{3}},\label{abpoly}
\end{eqnarray}
with $C_1$ being arbitrary constant which is a polynomial of $\xi_1$. Substituting (\ref{abpoly}) and $V_1=0$ into (\ref{abreduc2}), we can find the AB-KdV system (\ref{abkdv}) possesses a $P_s T_d$ symmetry breaking polynomial solution
\begin{eqnarray}
A=C_1\left(x-\frac{x_0}{2}\right)^3-6 C_1\left(t-\frac{t_0}{2}\right).
\end{eqnarray}
\textbf{Example 2.} If we take $m=0$, $n=-1$, then the solution of (\ref{abreduc11}) is
\begin{eqnarray}
&& U_1=C_2 \xi_1^{\frac{3}{2}} \Gamma\left(\frac{2}{3}\right) J_{-\frac{1}{3}}\left(\frac{2\sqrt{3}}{9}\xi_1^{\frac{3}{2}}\right)+\frac{2}{3}C_3 \pi \xi_1^{\frac{3}{2}} \Gamma\left(\frac{2}{3}\right)^{-1}J_{\frac{1}{3}}\left(\frac{2\sqrt{3}}{9}\xi_1^{\frac{3}{2}}\right),\nonumber \\ && \xi_1=\left(x-\frac{x_0}{2}\right)\left(t-\frac{t_0}{2}\right)^{-\frac{1}{3}},\label{abbessel}
\end{eqnarray}
where $\Gamma \left(\frac{2}{3}\right)$ is the indicated $\Gamma$ function and $J_{-\frac{1}{3}} \left(\frac{2 \sqrt{3}}{9} \xi_1^{\frac{3}{2}}\right)$, $J_{\frac{1}{3}}\left(\frac{2\sqrt{3}}{9}\xi_1^{\frac{3}{2}}\right)$ are the Bessel function, and $C_2$, $C_3$ are arbitrary constants. Substituting (\ref{abbessel}) and $V_1=0$ into (\ref{abreduc2}), we can find the AB-KdV system (\ref{abkdv}) has a Bessel function solution. Here we don't show it in detail.

Due to the rich and various solution structures of the reduction (\ref{abreduc1}) and arbitrary constants $m$ and $n$, we believe there are more possible reduction solutions of the AB-KdV system.
\section{B\"{a}cklund transformation of the AB-KdV system related to residual symmetry}
Residual symmetries have been successfully used to find some types of interaction solutions of nonlinear excitations. Further more, according to the results of the symmetry reductions with nonlocal symmetries, nonlocal residual symmetries can be used to find the Darboux transformations and B\"{a}cklund transformations for some nonlinear systems \cite{gaoloutang}. In this section, we discuss the residual nonlocal symmetries of the cKdV system (\ref{ckdv1}) - (\ref{ckdv2}), then localize the obtained nonlocal symmetries by introducing an enlarged system to provide the nonlocal symmetries related B\"{a}cklund transformation. At last, we will try to apply the known results to the AB-KdV system (\ref{abkdv}).
\subsection{B\"{a}cklund transformation theorem related to residual symmetry of the cKdV system}
The key point to find residual symmetries is to find the truncated Painlev\'{e} expansion of the system. For the cKdV equation (\ref{ckdv1})-(\ref{ckdv2}), its truncated Painlev\'{e} expansion has the form
\begin{eqnarray}
u=\frac{u_0}{f^2}+\frac{u_1}{f}+u_2,\qquad v=\frac{v_0}{f^2}+\frac{v_1}{f}+v_2, \label{pexpansion1}
\end{eqnarray}
where the function $f$ is the singularity manifold, and the functions $u_0$, $u_1$, $u_2$, $v_0$, $v_1$ and $v_2$ are related to the derivatives of $f$ and should be determined by substituting (\ref{pexpansion1}) into (\ref{ckdv1})-(\ref{ckdv2}) and vanishing the coefficients of $f^i$ for different $i$.

Substituting (\ref{pexpansion1}) into the cKdV equation (\ref{ckdv1}) - (\ref{ckdv2}), and vanishing the coefficients of $f^i$ for different $i$, we can obtain $u_0$, $u_1$, $u_2$, $v_0$, $v_1$ and $v_2$
\begin{eqnarray}
u_0=-\frac{1}{2}f_x^2,\nonumber
\end{eqnarray}
\begin{eqnarray}
u_1= \frac{1}{2}f_{xx} +q,\nonumber
\end{eqnarray}
\begin{eqnarray}
u_2=-\frac{1}{8}\frac{f_{xxx}}{f_x} +\frac{1}{16} \frac{f_{xx}^2}{f_x^2}+r-\frac{1}{24}\lambda, \nonumber
\end{eqnarray}
\begin{eqnarray}
v_0=-\frac{1}{2}f_x^2,\nonumber
\end{eqnarray}
\begin{eqnarray}
v_1=\frac{1}{2}f_{xx} -q,\nonumber
\end{eqnarray}
\begin{eqnarray}
v_2=-\frac{1}{8}\frac{f_{xxx}}{f_x} +\frac{1}{16} \frac{f_{xx}^2}{f_x^2}-r-\frac{1}{24}\lambda,\label{v2}
\end{eqnarray}
where $q$ and $r$ satisfy the following equations
\begin{eqnarray}
q_{xx}=-f_x r_x+\frac{f_{xx} q_x}{f_x}-\frac{1}{2}\frac{f_t}{f_x}q+\frac{1}{3}\lambda q, \nonumber
\end{eqnarray}
\begin{eqnarray}
q_t=2 f_x r_{xx}-2 f_{xx} r_x +\frac{2}{3} \lambda q_x -\frac{\lambda q f_{xx}}{3} +\frac{f_{xt} q}{f_x}, \nonumber
\end{eqnarray}
\begin{eqnarray}
r_t=-r_{xxx} +2 \lambda r_x-\frac{3}{2}\frac{f_t r_x} {f_x}+\frac{3}{2}\frac{f_{xx}^2 r_x}{f_x^2},\label{rt}
\end{eqnarray}
with $\lambda$ being an arbitrary constant and $f$ being a solution of the Shwarzian KdV equation \cite{louriccati}
\begin{eqnarray}
f_t=-f_{xxx}+\frac{3}{2}\frac{f_{xx}^2}{f_x}+\lambda f_x.\label{ft}
\end{eqnarray}

Now the truncated Painlev\'{e} expansion has been given. It can be proved that $\{u_1,\ v_1\}$ in the truncated Painlev\'{e} expansion (\ref{pexpansion1}) is a symmetry with respect to the solution $\{u_2, \ v_2\}$ of the cKdV system (\ref{ckdv1}) - (\ref{ckdv2}) which is called residue symmetry \cite{gaoloutang} related to Darboux transformations (DTs) and B\"{a}cklund transformations (BTs). Furthermore, the function $f$, $f_x$ and $f_{xx}$ is linked with $\{u, \ v\}$ non-locally via (\ref{v2}). The symmetry $\{u_1,\ v_1\}$ is nonlocal. We can verify it it true by substituting $\sigma^u=u_1$ and $\sigma^v=v_1$ with the solution $\{u_2, \ v_2\}$ of the cKdV system (\ref{ckdv1})-(\ref{ckdv2}) into the symmetry definition equation (\ref{susv}), where $q$, $r$ and $f$ satisfy the equations (\ref{rt}) - (\ref{ft}), with $\lambda$ being an arbitrary constant.

Since $\{u_1,\ v_1\}$ is a non-local symmetry related to DTs and BTs, one can naturally believe that it can be localized to a Lie point symmetry such that we can use the initial value problem to link two solutions of the cKdV system (\ref{ckdv1}) - (\ref{ckdv2}) due to the Lie's first principle. To localization the nonlocal symmetry $\{u_1,\ v_1\}$, we enlarge the cKdV system (\ref{ckdv1}) - (\ref{ckdv2}) into the following system
\begin{eqnarray}
u_t+u_{xxx}+3(u+v)(v_x+3u_x)=0,  \nonumber
\end{eqnarray}
\begin{eqnarray}
v_t+v_{xxx}+3(u+v)(u_x+3v_x)=0,\nonumber
\end{eqnarray}
\begin{eqnarray}
f_t=-f_{xxx}+\frac{3}{2}\frac{f_{xx}^2}{f_x}+\lambda f_x,\nonumber
\end{eqnarray}
\begin{eqnarray}
f_1=f_x,\nonumber
\end{eqnarray}
\begin{eqnarray}
f_2=f_{1x},\nonumber
\end{eqnarray}
\begin{eqnarray}
q_t=2 f_x r_{xx}-2 f_{xx} r_x +\frac{2}{3} \lambda q_x -\frac{\lambda q f_{xx}}{3} +\frac{f_{xt} q}{f_x}, \nonumber \label{qt1}
\end{eqnarray}
\begin{eqnarray}
q_{xx}=-f_x r_x+\frac{f_{xx} q_x}{f_x}-\frac{1}{2}\frac{f_t}{f_x}q+\frac{1}{3}\lambda q, \nonumber
\end{eqnarray}
\begin{eqnarray}
r_t=-r_{xxx} +2 \lambda r_x-\frac{3}{2}\frac{f_t r_x} {f_x}+\frac{3}{2}\frac{f_{xx}^2 r_x}{f_x^2},\nonumber
\end{eqnarray}
\begin{eqnarray}
u=-\frac{1}{8}\frac{f_{xxx}}{f_x} +\frac{1}{16} \frac{f_{xx}^2}{f_x^2}+r-\frac{1}{24}\lambda, \nonumber
\end{eqnarray}
\begin{eqnarray}
v=-\frac{1}{8}\frac{f_{xxx}}{f_x} +\frac{1}{16} \frac{f_{xx}^2}{f_x^2}-r-\frac{1}{24}\lambda,\nonumber
\end{eqnarray}
\begin{eqnarray}
&& s_{xx}=\frac{f_{xx} s_x}{ f_x}+\left(\frac{1}{2}\frac{f_{xxx}}{f_x} - \frac{3}{4}\frac{f_{xx}^2}{f_x^2}- \frac{1}{6} \lambda \right)s +4 f_x(fr_x-q_x). \label{sxx}
\end{eqnarray}
That means we have to solve the symmetry definition equations
\begin{eqnarray}
&&\sigma^u_t+\sigma^u_{xxx}+ 3 (\sigma^u+\sigma^v)(v_x+3 u_x)+3(u +v)(\sigma^v_x+3 \sigma^u_x)=0, \nonumber
\end{eqnarray}
\begin{eqnarray}
&& \sigma^v_t+\sigma^v_{xxx}+ 3 (\sigma^u+\sigma^v)(u_x+3v_x)+3(u+v)(\sigma^u_x+3\sigma^v_x)=0, \nonumber
\end{eqnarray}
\begin{eqnarray}
&& \sigma^f_x(f_x f_{xxx}+f_x f_t-3f_{xx}^2)-f_x(f_x \sigma^f_{xxx}+f_x \sigma^f_t-3f_{xx}\sigma^f_{xx})=0, \nonumber
\end{eqnarray}
\begin{eqnarray}
\sigma^{f_1}-\sigma^f_x=0, \nonumber
\end{eqnarray}
\begin{eqnarray}
\sigma^{f_2}-\sigma^f_{1x}=0, \nonumber
\end{eqnarray}
\begin{eqnarray}
&& 3 f_{xt} (f_x \sigma^q-\sigma^f_x q)+2f_{xx}(\lambda q \sigma^f_x-\lambda \sigma^q f_2-6 f_x^2 \sigma^r_x ) +12 f_x^2 r_{xx}\sigma^f_x+3 q f_x \sigma^f_{xt}\nonumber \\ && \qquad-2 \sigma^f_{xx}(\lambda q f_x+6r_x f_x^2)+12 f_x^3 \sigma^r_{xx}+2 f_x^2 (2\lambda \sigma^q_x-3 \sigma^q_t)=0, \nonumber
\end{eqnarray}
\begin{eqnarray}
&& 3 f_t q \sigma^f_x-12f_x^3 \sigma^r_x-2 f_x^2(\lambda q+6 \sigma^f_x r_x+3\sigma^q_{xx})-3f_x (\sigma^q f_t+ q \sigma^f-2f_{xx} \sigma^q_x\nonumber \\ && \qquad -2 q_x \sigma^f_{xx})-6 f_{xx} q_x \sigma^f_x=0, \nonumber
\end{eqnarray}
\begin{eqnarray}
&& f_x^3 (4 \lambda \sigma^r_x-2 \sigma^r_t-2 \sigma^r_{xxx})-3 f_x^2 (\sigma^r_x f_t+ \sigma^r_t r_x)+3 f_3 f_{xx}^2 \sigma+ 3 f_x r_t (f_t \sigma^f_x\nonumber \\ && \qquad +2 f_{xx} \sigma^f_{xx})-6 f_{xx}^2 r_x \sigma^f_x=0, \nonumber
\end{eqnarray}
\begin{eqnarray}
&& f_{xx}^2 \sigma^f_x+\sigma^f_{xxx} f_x^2 -f_x(\sigma^f_{xx} f_{xx}+\sigma^f_x f_{xxx})+8f_x^3 (\sigma^u -\sigma^r)=0, \nonumber
\end{eqnarray}
\begin{eqnarray}
&& f_{xx}^2 \sigma^f_x+\sigma^f_{xxx} f_x^2 -f_x(\sigma^f_{xx} f_{xx}+\sigma^f_x f_{xxx})+8f_x^3 (\sigma^v +\sigma^r)=0, \nonumber
\end{eqnarray}
\begin{eqnarray}
&& 6 \sigma^s(f_x f_{xxx}-3 f_{xx}^2)-12f_x \sigma^f (4 f_x^2 f r_x-4 f_x^2 q_x-s_x f_{xx}) -6 s(f_x\sigma^f_{xxx}\nonumber \\ && \qquad-3 f_{xx} \sigma^f_{xx})+\sigma^s f_x (2 \lambda f_x^2-6 f_x f_{xxx}+9 f_{xx}^2)-12 f_x^2 (4f f_x^2 \sigma^r_x\nonumber \\ && \qquad +4 f_x^2 r_x \sigma^f-4 f_x^2 \sigma^q_x-\sigma^s_{xx}f_x+\sigma^s_x f_{xx} +s_x \sigma^f_{xx})=0,
\end{eqnarray}
which are the linearized equations of the enlarged system (\ref{sxx}). It is not difficult to verify that the corresponding localized Lie point symmetry of (\ref{sxx}) has the form
\begin{eqnarray}
&& \sigma^u=\frac{1}{2}f_2+q, \qquad \sigma^v=\frac{1}{2}f_2-q, \qquad \sigma^f=-f^2, \nonumber \\ && \sigma^{f_1}=-2 f f_1, \qquad \sigma^{f_2}=-2 f f_2-2 f_1^2,\qquad \sigma^q=s, \nonumber \\ && \sigma^{s}= -3 s f, \qquad \sigma^r=q. \label{nonlockdv} \label{nonlocalsy}
\end{eqnarray}

The result (\ref{nonlocalsy}) tells us that the nonlocal symmetry (\ref{v2}) of the cKdV system (\ref{ckdv1}) - (\ref{ckdv2}) becomes a Lie point symmetry for the enlarged system (\ref{sxx}).

Whence a nonlocal symmetry is localized, it can be used to search for its finite transformations. By solving the initial value problem of the enlarged system (\ref{sxx})
\begin{eqnarray}
\frac{\textrm{d} u'(\epsilon)}{\textrm{d} \epsilon}&& =\frac{1}{2}f_2'(\epsilon)+q'(\epsilon), \qquad u'(0)=u, \nonumber
\\ \frac{\textrm{d} v'(\epsilon)}{\textrm{d} \epsilon}&& =\frac{1}{2}f_2'(\epsilon)-q'(\epsilon), \qquad v'(0)=v, \nonumber \\ \frac{\textrm{d} f'(\epsilon)}{\textrm{d} \epsilon}&& =-f'(\epsilon)^2, \qquad f'(0)=f, \nonumber \\ \frac{\textrm{d} f_1'(\epsilon)}{\textrm{d} \epsilon}&& =-2 f'(\epsilon) f_1'(\epsilon), \qquad f_1'(0)=f_1,\nonumber \\ \frac{\textrm{d} f_2'(\epsilon)}{\textrm{d} \epsilon}&& =-2 f'(\epsilon) f_2'(\epsilon)-2 f_1'(\epsilon)^2, \qquad f_2'(0)=f_2, \nonumber \\ \frac{\textrm{d} q'(\epsilon)}{\textrm{d} \epsilon}&& =s'(\epsilon), \qquad q'(0)=s, \nonumber \\ \frac{\textrm{d} s'(\epsilon)}{\textrm{d} \epsilon}&& = -3 s'(\epsilon) f'(\epsilon), \qquad s'(0)=s,\nonumber \\ \frac{\textrm{d} r'(\epsilon)}{\textrm{d} \epsilon}&&=q'(\epsilon), \qquad r'(0)=q,\label{inprob}
\end{eqnarray}
the B\"{a}cklund transformation theorem of the cKdV system (\ref{ckdv1}) - (\ref{ckdv2}) with finite transformations can be written down.

\textbf{Theorem 2} (B\"{a}cklund transformation theorem of the cKdV system (\ref{ckdv1}) - (\ref{ckdv2})). If $\{u, \ v, \ f, \ f_1, \ f_2, \ q, \ r, \ s \}$ is a solution of the enlarged system (\ref{sxx}), so $\{u', \ v', \ f', \ f_1', \ f_2', \ q', \ r', \ s' \}$ is a group invariant solution with
\begin{eqnarray}
u'&&=-\frac{\epsilon^2 f_1^2}{2(1+\epsilon f)^2}+\frac{\epsilon f_2}{2(1+\epsilon f)}+\frac{\epsilon^2 s}{2(1+\epsilon f)}+\epsilon q +u, \nonumber \\ v'&&=-\frac{\epsilon^2 f_1^2}{2(1+\epsilon f)^2}+\frac{\epsilon f_2}{2(1+\epsilon f)}-\frac{\epsilon^2 s}{2(1+\epsilon f)}-\epsilon q +v, \nonumber \\ f'&&=\frac{f}{1+\epsilon f}\nonumber \\ f_1'&&=\frac{f_1}{(1+\epsilon f)^2}, \nonumber \\ f_2'&&=\frac{f_2}{(1+\epsilon f)^2}-\frac{2 \epsilon f_1^2}{(1+\epsilon f)^3}, \nonumber \\ q' && =\frac{\epsilon (2+\epsilon f) s}{2(1+\epsilon f)^2}+q, \nonumber \\ s'&&=\frac{s}{(1+\epsilon f)^3}, \nonumber \\ r'&&=\frac{\epsilon^2 s}{2 (1+\epsilon f)}+ \epsilon q+r. \label{dtckdv}
\end{eqnarray}

According to theorem $2$, starting from any seed solutions of the cKdV system (\ref{ckdv1}) - (\ref{ckdv2}), we can obtain infinitely many new solutions, especially interaction solutions among different nonlinear excitations \cite{jincomm}.
\subsection{B\"{a}cklund transformation theorem related to residual symmetry of the AB-KdV system}
Based on the known results of the cKdV system (\ref{ckdv1}) - (\ref{ckdv2}), we now construct the B\"{a}cklund transformation for the AB-KdV system (\ref{abkdv}).

By applying the special reduction of $v=u(-x,-t)$ in the cKdV equation (\ref{ckdv2}), the solutions of the cKdV system (\ref{ckdv1}) - (\ref{ckdv2}) may be changed to those of the AB-KdV system (\ref{abkdv}) due to the $P_s T_d$ symmetric invariant. It is easy to verify the following B\"{a}cklund transformation theorem of the AB-KdV system (\ref{abkdv}) from the cKdV system.

\textbf{Theorem $3$} (B\"{a}cklund transformation theorem for AB-KdV system (\ref{abkdv})). If $A$ is solution of the AB-KdV system (\ref{abkdv}), $f$ is a group invariant solution of the Shwarzian KdV equation,
\begin{eqnarray}
f_t=-f_{xxx}+\frac{3}{2}\frac{f_{xx}^2}{f_x}+\lambda f_x, \label{ft2}
\end{eqnarray}
and $\{ s, \ q,\ r\}$ are anti-symmetric ($s(-x,\ -t)=-s(x,\ t)$, $q(-x,\ -t)=-q(x,\ t)$, $r(-x,\ -t)=-r(x,\ t)$) solutions of
\begin{eqnarray}
q_t=\frac{1}{2}\frac{f_{tx}}{f_x}-\left(2r_x+\frac{1}{3} \lambda \frac{ q}{f_x}\right)f_{xx} +2r_{xx}f_x + \frac{2}{3} \lambda q_x, \nonumber
\end{eqnarray}
\begin{eqnarray}
q_{xx}=\frac{1}{2} q \frac{f_{xxx}}{f_x} -\frac{3}{4} q \frac{f_{xx}}{f_x}^2+\frac{q_x f_{xx}}{f_x}-2r_x f_x -\frac{1}{6}\lambda q, \nonumber
\end{eqnarray}
\begin{eqnarray}
r_t=\frac{3}{2}\frac{r_x f_{xxx}}{f_x}-r_{xxx}- \frac{3}{4} \frac{r_x f_{xx}^2 }{f_x^2}+ \frac{1}{2} \lambda, \nonumber
\end{eqnarray}
\begin{eqnarray}
s_{xx}=\frac{f_{xx} s_x}{ f_x}+\left(\frac{1}{2}\frac{f_{xxx}}{f_x} - \frac{3}{4}\frac{f_{xx}^2}{f_x^2}- \frac{1}{6} \lambda \right)s+4 f_x(fr_x-q_x), \label{sxx2}
\end{eqnarray}
then
\begin{eqnarray}
A'(\epsilon)=-\frac{1}{2}\frac{\epsilon^2 f_x^2}{ (1+\epsilon f)^2}+ \frac{1}{2} \frac{\epsilon f_{xx}}{(1+\epsilon f)}+A,\label{dt1}
\end{eqnarray}
\begin{eqnarray}
&& A'(\epsilon)=-\frac{1}{2}\frac{\epsilon^2 f_x^2}{ (1+\epsilon f)^2}+ \frac{1}{2} \frac{\epsilon f_{xx}}{(1+\epsilon f)}+\frac{1}{2} \frac{\epsilon^2 s }{(1+\epsilon f)}+\epsilon q+A,\label{dt2}
\end{eqnarray}
are $P_sT_d$ invariant solution and $P_sT_d$ symmetry breaking solution of the AB-KdV system, respectively.

Theorem $3$ can be proved by direct substituting (\ref{dt1}) and (\ref{dt2}) into the AB-KdV system (\ref{abkdv}) using the Shwarzian KdV equation (\ref{ft2}) and the anti-symmetric solutions of (\ref{sxx2}).
\section{Summary and discussion}
In summary, a special AB-KdV system is directly obtained from KdV equation to describe two-place physical events by using $P_sT_d$ principle. Thus the AB-KdV system possessing $P_sT_d$ symmetry which means the AB-KdV system is invariant under the transformation $\{x\rightarrow -x+x_0,\ t\rightarrow -t+t_0\}$. Also, the AB-KdV system is nonlocal and can be used to describe two-place physical events.

With the help of a common KdV equation and a coupled KdV system, the exact solutions of the AB-KdV system are found, including $P_s T_d$ invariant and $P_s T_d$ symmetric breaking solutions with different methods. The $P_s T_d$ invariant solutions, such as multiple soliton solutions, soliton-cnoidal periodic wave interaction solutions and symmetry reduction solutions are obtained from those of the common KdV equation by the $P_s-T_d$ principle. The physical meaning of the group invariant solutions is that the event happened at $\{x,\ t\}$ will happen also at $\{-x+x_0,\ -t+t_0\}$. The $P_s T_d$ symmetric breaking solutions, such as single soliton solutions, singular soliton solutions and symmetry reduction solutions, are obtained by the help of a coupled KdV system which show rich structures of the AB-KdV system. The symmetry breaking solutions possesses the physical meaning that the event happened at $\{x,\ t\}$ will not happen also at $\{-x+x_0,\ -t+t_0\}$.

Also, a special B\"{a}cklund transformation related to residual symmetry is presented via the localization of the residual symmetry to find interactive solutions between the solitons and other types of the AB-KdV system.

In this manuscript, we find out the exact solutions by using the simply $P_s-T_d$ principle with the help of the common KdV equation and a coupled KdV equation. We believe our methods used here can be also applied in other AB systems to search for exact solutions. It can be seen from our results that AB systems possess rich structures which means the AB systems need further studies.

Because there exist various two-place (and multi-place) correlated physical events in almost all natural scientific fields, AB-physics/science may be widely and deeply affect all other scientific fields.

\section*{Acknowledgements}

The authors are indebt to thank Prof. Y. Chen and B. Li for their helpful discussions. The work is supported by NNSFC (Nos. 11675084 and 11435005) and Ningbo Natural Science Foundation (No. 2015A610159).  And the authors are sponsored by K. C. Wong Magna Fund in Ningbo University.
\nocite{*}

\bibliography{apssamp}
\bibliographystyle{elsarticle-num}

\end{document}